
\magnification=\magstep 1
\hsize=5 in
\vsize=6.75 in
\hoffset=.25 in
\voffset=.5 in

\def\tilde{\widetilde}
\def\bar{\overline}

\def\dsl{\raise.15ex\hbox{/}\kern-.57em\partial}
\def\Dsl{\,\raise.15ex\hbox{/}\mkern-13.5mu D}

\centerline{{\bf Deformation Analysis of Matrix Models}}
\vskip.1in\par\noindent
\centerline{\sl John Palmer }
\centerline{\sl Department of Mathematics }
\centerline{\sl University of Arizona }
\centerline{\sl Tucson, AZ 85721 }
\vskip.2in \par\noindent
{\bf Introduction}.  Suppose that $\varphi$ and $\psi$ are functions defined on
the real line and let
$$K(x,y)={{\varphi (x)\psi (y)-\psi (x)\varphi (y)}\over {x-y}}\chi
(y),\eqno (0.1)$$
where $\chi$ is the characteristic function of some union of
intervals
$$J=\cup_{j=1}^m[a_{2j-1},a_{2j}].$$
We regard $K(x,y)$ as the kernel of an integral operator $K$
$$Kf(x)=\int_{{\bf R}}K(x,y)f(y)\,dy.$$
For special choices of $\varphi$ and $\psi$, Fredholm determinants of the form
$$\hbox{det}(1-K),$$
arise in the theory of random Hermitian matrices [8].  In this theory
it is a result of Gaudin and Mehta that
such determinants give the probability that no eigenvalues
for an $N\times N$ random matrix lie in the set $J$ (the limit $N
\rightarrow\infty$ is
also of interest in the context of various scaling limits [12]).
More generally the derivative
$$\left.{{(-1)^n}\over {n!}}{{d^n}\over {d\lambda^n}}\hbox{det}(1
-\lambda K)\right|_{\lambda =1}$$
is the probability that exactly $n$ eigenvalues of the random
matrix lie in the interval $J$.

Fredholm determinants of a similar
sort arise for certain solvable models in two dimensional quantum
field theory [4], [3].  In the pioneering work [4], it was
shown for the simplest $N\rightarrow\infty$ limit in random matrix theory and
for
impenetrable Bosons in two dimensional field theory, that
it is possible to express the relevant Fredholm
determinant in terms of the solution to a completely integrable non-linear
system of
differential equations.  In [3] some other field theory models are
shown to have this same property.  In reference to this phenomenon,
Its, Izergin, Korepin, and Slavnov refer to
kernels of type (0.1) as ``completely integrable integral
kernels''.  Recently, Tracy and Widom [15], have given a systematic
treatment in the context of random matrix models.
They show that if the functions $\varphi$ and $\psi$ satisfy a linear
differential
equation
$${d\over {dz}}\left[\matrix{\varphi\cr
\psi\cr}
\right]=\Omega (z)\left[\matrix{\varphi\cr
\psi\cr}
\right],\eqno (0.2)$$
where
$$\Omega (z)={1\over {m(z)}}\left[\matrix{A(z)&B(z)\cr
-C(z)&-A(z)\cr}
\right],$$
and $m(z),A(z),B(z)$ and $C(z)$ are {\it polynomials}, then the Fredholm
determinant, $\hbox{det}(1-K)$, can be expressed in terms of the solution
to a system of non linear differential equations.

In this note
we will show that the differential equations found by Tracy and
Widom are special cases of monodromy preserving deformation
equations, and the analysis of the Fredholm determinant is a special
case of the general analysis of $\tau -\hbox{functions}$ [5].  This is not the
first
time that a connection between random matrices and monodromy
preserving deformation theory has been observed.  Moore, [9],
[10],
has already observed that families of differential equations of type
(0.2) play a role in the analysis of the partition function for
random matrix models.  In his work the singularity at infinity for
(0.2) is important.  The deformations of the coefficients $m,A,B$ and
$C$ that he considers all preserve the ``Stokes' multipliers'' for
(0.2), i.e., the monodromy data at infinity.  The partition function
is a $\tau -\hbox{function}$ for the resulting monodromy preserving
deformation,
and the general connection between $\tau -\hbox{functions}$ and non-linear
equations accounts, in particular, for the expression which gives
the partition function for two dimensional gravity in terms of a
Painlev\'e transcendent.  Some interesting mathematical problems
that arise in Moore's analysis have been solved by Beals and
Sattinger in [1].  In particular they prove that certain deformations
of (0.2), governed by compatible linear differential equations for
$\varphi$ and $\psi$,
do indeed leave the Stokes' multipliers at infinity invariant.

Here a local perturbation of (0.2) with simple poles at the points
$a_j$ will play a central role (see (1.14) below).  The compatibility of the
linear differential equations (1.14) and (1.15) will lead naturally to the
non-linear differential equations found by Tracy and Widom.
Realizing the non-linear equations as compatibility conditions for
linear systems is not new.  In [2] Harnad, Tracy
and Widom obtain the differential equations for the sine kernel
and the Airy kernel as compatibility conditions for linear systems
that are essentially the same as the linear systems we find below.
Apparently, these linear systems were guessed on the basis of
experience with the dynamical systems on coadjoint orbits of
loop algebras that are the central theme of [2].  One advantage of
our treatment is that the linear systems are very simple to get
at and the connection with dynamical systems on  coadjoint orbits
of loop algebras is then direct (in the setting which is explained
in [2]).

This paper is organized in two parts.  In the first part we show
how to realize the Tracy-Widom equations as monodromy preserving
deformation equations, staying close to the integral equation which is
central in their work.  In the second part we introduce the
formalism of Malgrange [7] in order to discuss the deformation
theoretic $\tau -\hbox{function}$.  Up to a trivial factor we show that this
is the determinant, $\hbox{det}(1-\lambda K)$.  The formalism of Malgrange is,
incidently, better adapted to deal with analyticity questions in the $
a_j$
parameters.

The author would like to thank C.~Tracy and H.~Widom for helpful
conversations about their work.
\vfill\eject

\centerline{\S 1 {\bf The Deformation Problem}}
\vskip.1in\par\noindent
We will now introduce the deformation problem which will lead to
the linear equations (1.14) and (1.15).  We will consider the problem in
light of the formalism for Cauchy-Riemann operators
introduced in [11].  It is not essential to understand the formalism
in [11] to follow what goes on but it does provide some motivation.
Suppose to begin that
$$\left[\matrix{\varphi (z)\cr
\psi (z)\cr}
\right]$$
is a solution to the differential equation (0.2) defined for $z\in
{\bf C}$, the
complex plane.  Here $m,A,B,$ and $C$ are assumed to be polynomials
as mentioned above.  In case $m$ is a non constant polynomial it
may happen that the differential equation (0.2) does not have any
globally defined single valued solutions.  To avoid this complication at the
start
we will suppose that $m=1$.
Let
$$\left[\matrix{\tilde{\varphi }(z)\cr
\tilde{\psi }(z)\cr}
\right]$$
be a second independent solution to the same differential
equation (0.2).  Since the trace of the matrix on the right hand side
of (0.2) is 0 the determinant of the fundamental solution
$$\left[\matrix{\varphi&\tilde{\varphi}\cr
\psi&\tilde{\psi}\cr}
\right]$$
is constant and we can choose $\tilde{\varphi}$ and $\tilde{\psi}$ so that this
determinant is 1.   We wish to consider the Cauchy Riemann
operator
$$\left[\matrix{\varphi&\tilde{\varphi}\cr
\psi&\tilde{\psi}\cr}
\right]^{-1}\left[\matrix{\bar{\partial}_z&0\cr
0&\bar{\partial}_z\cr}
\right]\left[\matrix{\varphi&\tilde{\varphi}\cr
\psi&\tilde{\psi}\cr}
\right].\eqno (1.1)$$
In this formula $ $ $\bar{\partial}_z={1\over 2}(\partial_x+i\partial_
y)$ as usual.
Of course, since $\varphi$, $\tilde{\varphi}$, $\psi$, and $\tilde{
\psi}$ are analytic
functions in {\bf C} this C-R operator acts precisely as the standard
C-R operator
$$\left[\matrix{\bar{\partial}_z&0\cr
0&\bar{\partial}_z\cr}
\right].\eqno (1.2)$$
The difference is that we will take the domain of (1.2) to be
the Sobolev space $H^1({\bf C})$ so that (1.1) represents an operator with
with domain
$$\left[\matrix{\varphi&\tilde{\varphi}\cr
\psi&\tilde{\psi}\cr}
\right]^{-1}H^1({\bf C}).$$
Note that since the determinant of the fundamental solution is
1 the inverse
$$\left[\matrix{\varphi&\tilde{\varphi}\cr
\psi&\tilde{\psi}\cr}
\right]^{-1}=\left[\matrix{\tilde{\psi}&-\tilde{\varphi}\cr
-\psi&\varphi\cr}
\right].$$
Next we wish to introduce a different domain for the C-R
operator (1.1) which incorporates functions with fixed branching
behavior along an interval $[a_1,a_2]$ on the real axis.  It will be obvious
how to incorporate more than one interval $[a_j,a_{j+1}]$ in the formalism
so I will confine my attention to just one interval for simplicity.
We wish to
include in the domain of our new C-R operator functions which
change by a factor of
$$M=\left[\matrix{1&2\pi i\lambda\cr
0&1\cr}
\right]$$
in a counterclockwise circuit of the left endpoint, $a_1,$ and by
the inverse of this matrix in a counterclockwise circuit of the
right endpoint, $a_2$.  To understand the resulting C-R operator
we would like to realize it as a similarity transform
$$Y^{-1}\left[\matrix{\bar{\partial}_z&0\cr
0&\bar{\partial}_z\cr}
\right]Y$$
of the standard C-R operator by a multivalued analytic function
$Y(z)$.  So that the domain of this operator is the same as the
domain of (1.1) near infinity we want
$$Y(z)\sim F(z)\hbox{  for }z\hbox{  near  }\infty ,\eqno (1.3)$$
where we write
$$F(z)=\left[\matrix{\varphi (z)&\tilde{\varphi }(z)\cr
\psi (z)&\tilde{\psi }(z)\cr}
\right].$$
To find $Y(z)$ we split the problem into two pieces.  Let
$Y_{+}(z)=Y(z)$ for $z\in \{z:z_2\ge 0\}$ and $Y_{-}(z)=Y(z)$ for $
z\in \{z:z_2\le 0\}.$
Then a more precise formulation of (1.3) is that
$$FY_{\pm}^{-1}-\left[\matrix{1&0\cr
0&1\cr}
\right]\in H_{\pm},\eqno (1.4)$$
where $H_{\pm}$ is the usual Hardy space for the upper $(+)$ and lower
(-) half planes.  On the real axis the boundary values $Y_{+}$ and $
Y_{-}$
are related by
$$Y_{+}=Y_{-}\left((I-\chi )+\chi M\right)$$
where $\chi =\chi_{[a_1,a_2]}$ is the characteristic function of the interval.
In order that the integral operator $K$ (rather than its transpose)
arises in the fundamental integral equation (1.6) below it will be
convenient to temporarily work with $Y_{\pm}^{-1}$.  The monodromy condition
relating the boundary values of $Y_{\pm}^{-1}$ becomes
$$Y_{-}^{-1}=\left(\left(I-\chi\right)+\chi M\right)Y_{+}^{-1}.$$
We can rewrite this last equation as
$$Y_{-}^{-1}-F^{-1}=Y_{+}^{-1}-F^{-1}+\chi (M-I)Y_{+}^{-1}.\eqno
(1.5)$$
Let $P_{+}$ be the usual orthogonal projection on $H_{+}$ and define
$$P^F_{+}=F^{-1}P_{+}F.$$
Then $P_{+}^F$ is the analogue of $P_{+}$ for the C-R operator (1.1).
We can write (1.4) as
$$\eqalign{P_{+}^F(Y_{-}^{-1}-F^{-1})&=0,\cr
P_{+}^F(Y_{+}^{-1}-F^{-1})&=Y_{+}^{-1}-F^{-1},\cr}
$$
where we are not multiplying operators anymore but just
applying $P_{+}^F$ to the columns of the matrices $Y_{\pm}^{-1}-F^{
-1}.$
Applying $P_{+}^F$ to the columns of (1.5) and making use of these
last relations one finds

$$Y_{+}^{-1}+P_{+}^F\chi (M-I)Y_{+}^{-1}=F^{-1}.\eqno (1.6)$$
Since the kernel of $P_{+}$ is $-{1\over {2\pi i}}(x-y)^{-1}$ (with $
x=x+i0$)
we find that the kernel of the operator
$P_{+}^F\chi (M-I)$ is
$$-{{\lambda\chi (y)}\over {x-y}}\left[\matrix{0&\tilde{\psi }(x)
\varphi (y)-\tilde{\varphi }(x)\psi (y)\cr
0&\varphi (x)\psi (y)-\psi (x)\varphi (y)\cr}
\right].$$
Now let $K$ denote the operator with kernel
$${{\chi (y)}\over {x-y}}\left(\varphi (x)\psi (y)-\psi (x)\varphi
(y)\right).$$
This is a non-singular kernel.  Let $\tilde {K}$ denote the operator with
kernel
$${{\chi (y)}\over {x-y}}\left(\tilde\psi (x)\varphi (y)-\tilde\varphi
(x)\psi (y)\right).$$
This is a singular kernel.  Because of the connection with $P_{+}$ it is
understood that the variable $x$ should be allowed to
approach the real axis from above (i.e., $x+i0$).  In terms of $K$ and
$\tilde {K}$ the equation (1.6) becomes
$$\left[\matrix{1&-\lambda\tilde {K}\cr
0&1-\lambda K\cr}
\right]Y_{+}^{-1}=F^{-1}.\eqno (1.7)$$
This is solvable provided $1-\lambda K$ is invertible.  We suppose that
this is the case in what follows.  As we will see in the second
section, the $\tau -\hbox{function}$ for this problem is just $\hbox{det}
(1-\lambda K)$.

Solving equation (1.7) for $Y_{+}^{-1}$ one finds
$$Y_{+}^{-1}=\left[\matrix{\tilde{\psi }-\lambda\tilde {K}P&-\tilde{
\varphi }+\lambda\tilde {K}Q\cr
-P&Q\cr}
\right],$$
where
$$\eqalign{Q=&(1-\lambda K)^{-1}\varphi ,\cr
P=&(1-\lambda K)^{-1}\psi .\cr}
$$
We have here adopted the notation from [15] to facilitate comparisons.
{}From the integral equation
$$FY_{+}^{-1}=I-P_{+}\left(F\chi (M-I)Y_{+}^{-1}\right),\eqno (1.
8)$$
it is clear that $FY_{+}^{-1}$ has an analytic continuation into the upper
half plane.  Since $F$ is analytic and invertible in the whole
plane, $Y_{+}$ will be analytic in the upper half plane.  The same argument
shows that $Y_{-}$ has an analytic continuation into the lower
half plane.  The two functions $Y_{+}$ and $Y_{-}$ fit together to give
a single analytic function $Y(z)$ which has a branch cut joining $
a_1$ and
$a_2$.  We want to analyse the singularities of $Y(z)$ in a
neighborhood of the points $a_j$ and near $\infty$.  First we consider
the local singularity at $a_1$.  Define $f(z)=\log(z-a_1)$, with the branch
cut for the logarithm taken to be the ray $[a_1,\infty )$.  Apart
from the location of the branch cut it does not matter which
determination of the logarithm we use.
We claim that the matrix
$$Y(z)\left[\matrix{1&\lambda f(z)\cr
0&1\cr}
\right]=G(z),\eqno (1.9)$$
is single valued and analytic in a neighborhood of $z=a_1$.  The
left hand side of (1.9) is single valued because the monodromy
of the matrix
$$\left[\matrix{1&\lambda f(z)\cr
0&1\cr}
\right],$$
as $z$ makes a circuit of $a_1$ exactly cancels the monodromy that
is built into $Y(z)$ by construction.  The left hand side of (1.9) will
have a removable singularity at $a_1$ because
 $P(z)$ and $Q(z)$ are regular
at $z=a_1$ and $\tilde {K}$  introduces only a logarithmic singularity at $
z=a_1$.

Taking determinants of both sides of (1.9) shows that $\hbox{det}
(Y(z))$ is
an analytic function of $z$ near $z=a_1$.  In a similar fashion one
can show that $\hbox{det}(Y(z))$ is analytic near $z=a_2$.  From (1.8) it
follows that
$$\hbox{det}(F^{-1}Y(z))=\hbox{det}(Y(z))=1+O(z^{-1}),$$
for $z$ in a neighborhood of $\infty .$  Thus $\hbox{det}(Y(z))=1$ for all $
z$.  It
follows from this and the earlier expression for $Y$ that
$$Y=\left[\matrix{Q&\tilde{\varphi }-\lambda\tilde {K}Q&\cr
P&\tilde{\psi }-\lambda\tilde {K}P\cr}
\right],$$
which displays the analogous roles played by $\varphi$, and $\psi$ and
$Q$, and $P$.

Differentiating (1.9) with respect to $z$ one finds
$${{\lambda Y(z)}\over {z-a_1}}\left[\matrix{0&1\cr
0&0\cr}
\right]+Y'(z)\left[\matrix{1&\lambda f(z)\cr
0&1\cr}
\right]=G'(z),$$
which may be easily rewritten
$$Y'(z)Y^{-1}(z)=-{{\lambda}\over {z-a_1}}Y(z)\left[\matrix{0&1\cr
0&0\cr}
\right]Y^{-1}(z)+G'(z)G^{-1}(z).\eqno (1.10)$$
Now using the fact that the determinant of $Y$ is 1 and the
formula for $Y$ in terms of $P$ and $Q$ above one finds that
$$Y'(z)Y^{-1}(z)={{\lambda}\over {z-a_1}}\left[\matrix{q_1p_1&-q_
1^2\cr
p_1^2&-q_1p_1\cr}
\right]+O(1)\hbox{  for  }z\sim a_1.\eqno (1.11)$$
In this formula we've written $q_j=Q(a_j)$ and $p_j=P_{}(a_j).$  In a
precisely analogous fashion we calculate the residue of $Y'Y^{-1}$
at $z=a_2$ to find
$$Y'(z)Y^{-1}(z)=-{{\lambda}\over {z-a_2}}\left[\matrix{q_2p_2&-q_
2^2\cr
p_2^2&-q_2p_2\cr}
\right]+O(1)\hbox{  for  }z\sim a_2.\eqno (1.12)$$
Now define
$$\Delta =-P_{+}F(M-I)\chi Y_{+}^{-1}.$$
Differentiating (1.8) with respect to $z$ one finds
$$F'Y^{-1}-FY^{-1}Y'Y^{-1}=\Delta',$$
or
$$F'F^{-1}(FY^{-1})-(FY^{-1})Y'Y^{-1}=\Delta'.$$
Since $FY^{-1}=I+\Delta$ this can be written
$$Y'Y^{-1}=(I+\Delta )^{-1}\Omega (I+\Delta )-(I+\Delta )^{-1}\Delta'
.$$
Since $\Delta (z)=O(z^{-1})$ and $\Delta'(z)=O(z^{-2})$ it follows that the
principal part of $Y'Y^{-1}$ at $\infty$ is the same as the principal
part of
$$(I+\Delta )^{-1}\Omega (I+\Delta ).\eqno (1.13)$$
It now follows from this observation and (1.11) and (1.12) that there
is a matrix valued polynomial $M(z)$ of the same degree as
$\Omega (z)$ so that
$${{dY}\over {dz}}Y^{-1}=M(z)+{{\lambda C_1}\over {z-a_1}}-{{\lambda
C_2}\over {z-a_2}},\eqno (1.14)$$
where
$$C_k=\left[\matrix{q_kp_k&-q_k^2\cr
p_k^2&-q_kp_k\cr}
\right].$$

Now we determine the differential equation satisfied by $Y$ in the
$a_j$ variables.  Differentiating (1.8) in the $a$ variables one finds
$$d_aYY^{-1}=O(z^{-1})\hbox{  as  }z\rightarrow\infty .$$
Differentiating (1.9) in the $a$ variables one finds the principal part of
$d_aYY^{-1}$ at $a_j$ is:
$$(-1)^k{{\lambda C_k}\over {z-a_k}}da_k.$$
Combining these two observations one finds that the function
$d_aYY^{-1}$ is analytic in $z$ on ${\bf P}^1$ except for simple poles at $
a_1$ and
$a_2$.  It has a zero at $z=\infty$ and so
$$d_aYY^{-1}=-{{\lambda C_1}\over {z-a_1}}da_1+{{\lambda C_2}\over {
z-a_2}}da_2.\eqno (1.15)$$
Now we apply these observations to deduce non-linear deformation
equations.  If one examines the equality of mixed partials
$${{\partial}\over {\partial a_j}}{{\partial}\over {\partial z}}Y
={{\partial}\over {\partial z}}{{\partial}\over {\partial a_j}}Y,$$
using (1.14) and (1.15), one obtains an equation relating rational
functions of $z$ with poles at $a_j$ and $\infty$.  Equating the principal
parts
at the finite poles $a_j$, one finds
$$\eqalign{d_aC_1&=[M(a_1),C_1]da_1+\lambda [C_1,C_2]{{d(a_2-a_1)}\over {
a_2-a_1}},\cr
d_aC_2&=[M(a_2),C_2]da_2+\lambda [C_1,C_2]{{d(a_2-a_1)}\over {a_2
-a_1}},\cr}
\eqno (1.16)$$
where $M(a_j)$ is the value of the matrix polynomial $M(z)$ at
$z=a_j$.  Equating the principal parts at $\infty$ one finds
$${{\partial M(z)}\over {\partial a_j}}=(-1)^j\lambda\left[{{M(z)
-M(a_j)}\over {z-a_j}},C_j\right].\eqno (1.17)$$
Note that in this last equation each side is a polynomial function
of $z$ of order 1 less than the degree, $n,$ of $M(z)$ (since the highest
order term in $M(z)$ is clearly independent of $a_j$).  In particular if
we write
$$M(z)=M_nz^n+M_{n-1}z^{n-1}+\ldots$$
then it is easy to see that (1.17) implies
$${{\partial M_{n-1}}\over {\partial a_j}}=(-1)^j\lambda\left[M_n
,C_j\right].$$
In order to make (1.16) and (1.17) more explicit we must use (1.13) to
compute the polynomial $M(z)$.
To make use of (1.13) we must in turn calculate the asymptotics of $
\Delta (z)$
at $z=\infty$.
Substituting for $Y_{+}$ and $F$ in the formula for $\Delta (z)$ one finds
$$\Delta (z)=\int_{{\bf R}}{{\lambda\chi (y)}\over {z-y}}\left[\matrix{
-\varphi (y)P(y)&\varphi (y)Q(y)\cr
-\psi (y)P(y)&\psi (y)Q(y)\cr}
\right]\,dy.$$
Expanding the denominator in a geometric series
$${1\over {z-y}}={1\over z}\left(1+{y\over z}+\left({y\over z}\right
)^2+\cdots\right)$$
one finds that
$$\Delta (z)={{\lambda}\over z}\left[\matrix{-\tilde {v}_0&u_0\cr
-w_0&v_0\cr}
\right]+{{\lambda}\over {z^2}}\left[\matrix{-\tilde {v}_1&u_1\cr
-w_1&v_1\cr}
\right]+O(z^{-3}),\eqno (1.18)$$
where
$$\eqalign{u_j&=\int_{{\bf R}}\chi (y)y^j\varphi (y)Q(y)\,dy,\cr
\tilde {v}_j&=\int_{{\bf R}}\chi (y)y^j\varphi (y)P(y)\,dy,\cr
v_j&=\int_{{\bf R}}\chi (y)y^j\psi (y)Q(y)\,dy,\cr
w_j&=\int_{{\bf R}}\chi (y)y^j\psi (y)P(y)\,dy,\cr}
\eqno (1.19)$$
and the notation is again taken from [15].  We use (1.18) and (1.13) to
calculate the matrix $M(z)$ in a few of the cases considered in
[15].

Our first example is the {\it sine} kernel with
$$\eqalign{\varphi (z)&=\sin(z),\cr
\psi (z)&=\cos(z).\cr}
$$
In this case
$$\Omega (z)=\left[\matrix{0&1\cr
-1&0\cr}
\right],$$
and a simple calculation shows that
$$M_{\hbox{Sine}}(z)=\left[\matrix{0&1\cr
-1&0\cr}
\right].\eqno (1.20)$$

Our second example is the {\it Airy} kernel with
$$\eqalign{\varphi (z)&=\hbox{Ai}(z),\cr
\psi (z)&=\hbox{Ai}'(z).\cr}
$$
In this case
$$\Omega (z)=\left[\matrix{0&1\cr
z&0\cr}
\right].$$
Without difficulty one may use (1.13) and (1.18) to compute
$$M_{\hbox{Airy}}(z)=\left[\matrix{-\lambda u&1\cr
z-\lambda (v+\tilde {v})&\lambda u\cr}
\right],\eqno (1.21)$$
where we've written $u=u_0$, $v=v_0$ and $\tilde {v}=\tilde {v}_0$.  For $
\lambda =1$ this
simplifies somewhat since $v_0=\tilde {v}_0$ in this event.

Our last example is one that arises in 2D quantum {\it gravity} matrix
models (the case of pure gravity).  A function $\xi (x)$ is given which
satisfies the differential equation
$$\xi^{\prime\prime\prime}(x)+6\xi (x)\xi'(x)+4=0,$$
which is a special case of the string equation.
The functions $\varphi (z,x)$ and $\psi (z,x)$ now depend on the parameter $
x$ and
satisfy the differential equation
$${d\over {dz}}\left[\matrix{\varphi\cr
\psi\cr}
\right]=\Omega (z)\left[\matrix{\varphi\cr
\psi\cr}
\right],$$
where
$$\Omega (z)=\left[\matrix{-{1\over 4}\xi'(x)&z+{1\over 2}\xi (x)\cr
-z^2-{1\over 2}\xi (x)z-{1\over 2}\xi^2(x)-{1\over 4}\xi^{\prime\prime}
(x)&{1\over 4}\xi'(x)\cr}
\right].$$
This is incidently, an example of a family of differential equations
in which the Stokes' multipliers at $\infty$ are independent of $
x$
[9], [10].
Again one may use (1.13) and (1.18) to compute
$$M_{\hbox{Gravity}}(z)=\Omega (z)+M_1z+M_0$$
where
$$M_1=\left[\matrix{\lambda u&1\cr
\lambda (v+\tilde {v})&-\lambda u\cr}
\right]$$
and
$$M_0=\lambda\left[\matrix{-\lambda uv+{1\over 2}\xi (x)u-w+u_1&\lambda
u^2+(v+\tilde {v})\cr
\lambda\left(uw-v(v+\tilde v)\right)+\xi (x)v+\tilde {v}_1+v_1&\lambda
uv-{1\over 2}\xi (x)u+w-u_1\cr}
\right]$$
where we've written $u=u_0$, $v=v_0$ and $\tilde {v}=\tilde {v}_0$.

It is straightforward to use these results to determine the differential
equations (1.16) and (1.17).  We will not do this here but instead we
will refer the reader to [15], where the non-linear equations for $
q_j$, $p_j,$
$u_j,$ $\ldots$ are written out, and some interesting examples beyond
those mentioned above are considered.  Tracy and
Widom also show that equations (1.16) and (1.17) sometimes have
additional first integrals which can be used to reduce the number
of variables in the
deformation equations.  In several examples they show that
the the deformation equations can be integrated in terms of
Painlev\'e transcendents (thus generalizing the original observation in
[4] that Painlev\'e V transcendents arise in the integration of the
deformation equations associated with the sine kernel).

To conclude this section we will describe the
changes which need to be made when the differential equation
for $\varphi$ and $\psi$ already has poles in the finite plane.  The simplest
example of this sort considered by Tracy and Widom is the
{\it Bessel} kernel, for which
$$\eqalign{\varphi (z)&=J_{\alpha}(\sqrt z),\cr
\psi (z)&=zJ'_{\alpha}(\sqrt z).\cr}
$$
One can see from this that $\varphi (z)$ and $\psi (z)$ will not be single
valued in the complex plane.  One can deal with this situation by
introducing a branch cut for $\varphi$ and $\psi$ joining 0 to $\infty$.  Since
the
intervals $[a_j,a_{j+1}]$ in this model will all be subsets of the positive
real line it is convenient, in this case, to choose the branch cut on the
negative
real axis.  The differential equation satisfied by $\varphi$ and $
\psi$ is
$${d\over {dz}}\left[\matrix{\varphi\cr
\psi\cr}
\right]={1\over z}\left[\matrix{0&1\cr
{1\over 4}\left(\alpha^2-z\right)&0\cr}
\right]\left[\matrix{\varphi\cr
\psi\cr}
\right],$$
which has a simple pole at 0 in addition to an irregular singular
point at infinity.  Extending $\left[\matrix{\varphi\cr
\psi\cr}
\right]$ to a matrix valued fundamental
solution $F$ with determinant 1 is possible as before provided one
admits a branch cut for $F$ on the negative real axis.  Let $F_{+}$
denote the analytic function $F$ in the upper half plane and let
$F_{-}$ denote the analytic function $F$ in the lower half plane.  The
boundary values of these two functions on the real axis are
related by
$$F_{+}=F_{-}\left(I+\chi_0(M_0-I)\right),\eqno (1.22)$$
where $\chi_0$ is the characteristic function of the interval $(-
\infty ,0]$
and $M_0$ is the monodromy matrix associated with the
fundamental solution $F$.  We will want the monodromy of the
fundamental solution $Y$ to match that of $F$ on the negative real
real axis and so the relation we desire between the boundary
values $Y_{\pm}$ is
$$Y_{-}^{-1}=\left(I+\chi_0(M_0-I)+\chi (M-I)\right)Y_{+}^{-1}\eqno
(1.23)$$
Using (1.22) to write $F_{-}$ in terms of $F_{+}$, and the fact that
$\chi_0\chi =0$, one can multiply (1.23) on the left by $F_{-}$ to get
$$F_{-}Y_{-}^{-1}-I=F_{+}Y_{+}^{-1}-I+F_{+}\chi (M-I)Y_{+}^{-1}.$$
If we now insist that
$$P_{\pm}\left(F_{\pm}Y_{\pm}^{-1}-I\right)=F_{\pm}Y_{\pm}^{-1}-I
,$$
as before, then we find
$$F_{+}Y_{+}^{-1}=I-P_{+}F_{+}(M-I)\chi Y_{+}^{-1},\eqno (1.24)$$
instead of (1.8).  Now let
$$\Delta =-P_{+}F(M-I)\chi Y_{+}^{-1}$$
and differentiate (1.24) with respect to $z$ to obtain
$$Y_{+}'Y_{+}^{-1}=(I+\Delta )^{-1}\Omega (I+\Delta )-(I+\Delta )^{
-1}\Delta'.$$
Now since $\Delta (z)=O(z^{-1})$ and $\Delta'(z)=O(z^{-2})$ we can conclude
that
the principal part of $zY_{+}'(z)$ $Y_{+}(z)^{-1}$ at $z=\infty$ is the same as
the
principal part of
$$(I+\Delta (z))^{-1}z\Omega (z)(I+\Delta (z)).\eqno (1.25)$$
The calculation of the principal part of (1.25) at infinity
simultaneously determines the residues for $Y'Y^{-1}$ at $\infty$ and at
0.  The singularity analysis of $Y'Y^{-1}$ near $a_j$ for $j=1,2$ is
unchanged from before and one finds
$$Y'Y^{-1}=-{1\over 4}\left[\matrix{0&0\cr
1&0\cr}
\right]+{{\lambda}\over {4z}}\left[\matrix{u&4\cr
\alpha^2+v+\tilde {v}&-u\cr}
\right]+{{\lambda C_1}\over {z-a_1}}-{{\lambda C_2}\over {z-a_2}}
,\eqno (1.26)$$
where we've written $u=u_0$, $v=v_0$ and $\tilde {v}=\tilde {v}_0$.  The
analysis
of the singularities in $d_aYY^{-1}$ proceeds as above and we find
(1.15) is unchanged.  The non-linear equations are deduced as
compatibility conditions for (1.26) and (1.15).  Once again we
refer the reader to [14] and [15] for a more detailed look at the
resulting equations.

\vskip.2in \par\noindent

\centerline{\S 2 {\bf The tau function}}
\vskip.1in\par\noindent
In this section we will show that the determinant $\hbox{det}(1-\lambda
K)$ which is
the focal point of the random matrix applications is the $\tau
-\hbox{function}$
for a deformation problem.  As in the first section there is no
serious loss of generality in considering the case of just a single
interval $(a_1,a_2)$ and so for simplicity we shall do so.  It is also
simpler to discuss the case in which $m(z)=1$, and although it is
not essential to do so, we will confine our attention in \S 2 to
this special case.

We begin by
reminding the reader of a formula for the logarithmic derivative
of $\hbox{det}(1-\lambda K)$ which can be found in [15],
$$d_a\log\hbox{det}(1-\lambda K)=R(a_1,a_1)da_1-R(a_2,a_2)da_2.\eqno
(2.0)$$
In this formula the terms $R(a_j,a_j)$ are the diagonal values for the
kernel of the resolvent operator, $\lambda K(1-\lambda K)^{-1}$, for $
\lambda K$.  In [15] it is
shown that
$$R(a_k,a_k)=p_k{{\partial q_k}\over {\partial a_k}}-q_k{{\partial
p_k}\over {\partial a_k}},\eqno (2.1)$$
where the notation
$$q_k=Q(a_k),\hbox{  }p_k=P(a_k),$$
is as in $\S 1$.  We write
$$M(z)=\left[\matrix{M_{11}(z)&M_{12}(z)\cr
M_{21}(z)&M_{22}(z)\cr}
\right]$$
in (1.16) and find, using (2.1), that,
$$R(a_k,a_k)=\hbox{Tr}(M(a_k)C_k)+(-1)^{k-1}\lambda{{\hbox{Tr}(C_
1C_2)}\over {a_2-a_1}}.\eqno (2.2)$$
We write Tr to denote the {\it trace} of a matrix.
The notation $M(a_j)$ is slightly misleading since $M(z)$ is a
polynomial in $z$ with coefficients that are functions of
$a=(a_1,a_2).$  To make the dependence on the point $a$ a little clearer
we should write $M(z,a)$ for $M(z)$.  Then $M(a_j)=M(a_j,a).$
We will use both notations depending on circumstances.

Using (2.0) and (2.2) we will make contact with the
formula for the log derivative of the tau function.  In [5] Jimbo,
Miwa and Ueno have given a formula for the log derivative of the
tau function in the general case of irregular singular points.  In
[7] Malgrange has provided a geometrical significance for the tau
function in the case of regular singular points with simple poles.
Instead of simply adopting
the JMU formula we will introduce the slight modifications in
Malgrange's analysis that are necessary to deal with the irregular
singular point at infinity in (1.14).  We hope that this will prove to
be more instructive.

First, fix points $a_j^0\in {\bf R}\subset {\bf C}$ which are distinct
for $j=1,2.$  We suppose that the integral operator $1-K_0$ is
invertible.  Here $K_0$ is the integral operator with kernel given by
(0.1) and the interval $J=(a_1^0,a_2^0).$  Since $1-K_0$ is invertible and
$K_0$ is compact it is clear that $1-\lambda K_0$ will remain invertible
provided that $\lambda\in {\bf C}$ is chosen in a sufficient small neighborhood
of 1.  Throughout this section we will suppose that $\lambda$ is
chosen sufficiently close to 1.

We may rephrase the existence
of $Y(z,a^0)$ satisfying (1.14) in the following terms.  There exists a
holomorphic connection $\nabla^0$ on the trivial bundle ${\bf C}\backslash
\{a_1^0,a_2^0\}\times {\bf C}^2$ with
simple poles at $a_1^0$ and $a_2^0$ given by,
$$\nabla^0=d_z-\Omega (z,a^0)dz,$$
where,
$$\Omega (z,a)=M(z,a)+{{\lambda C_1(a)}\over {z-a_1}}-{{\lambda C_
2(a)}\over {z-a_2}},$$
and,
$$d_z=dz\partial_z+d\bar {z}\bar{\partial}_z.$$
Note that $\nabla^0$ is an integrable connection
(it has curvature 0) since
$${{\partial}\over {\partial\bar {z}}}\Omega (z,a^0)=0$$
for $z\in {\bf C}\backslash \{a_1^0,a_2^0\},$ and locally flat sections for $
\nabla^0$ are ${\bf C}^2$ valued
holomorphic functions, $s(z),$ satisfying the differential equation
$$s'(z)=\Omega (z,a^0)s(z).$$
One problem considered by Malgrange in [7] is the problem of
deforming the connection $\nabla^0$ in the ``$a$'' variables so that the
resulting connection, $\nabla ,$  has simple poles at $a_j$ for $
j=1,2$ and so
that the holonomy of the resulting connection about the
singular points $a_j$ remains fixed as the points $a_j$ vary.
Actually, to fix the
determination of holonomy for a connection we must first
choose a base point, $z_0$, and closed simple paths, $\gamma_j,$ about the
singular points $a_j$.  Parallel translation around the curves $\gamma_
j$
then depends only on the homotopy class of the path $\gamma_j$ in
${\bf C}\backslash \{a_1,a_2\}$.   Parallel
translation about a closed curve is given by a linear
transformation on the fiber over the endpoint which our
fixed trivialization identifies with ${\bf C}^2$.  One might consider the
problem of
deforming the connection $\nabla^0$ while keeping the holonomy matrices
associated with the curves $\gamma_j$ fixed.  However, to make contact with the
work we did in the first section this is not exactly the
problem we will consider.
Our principal difficulty is that we would like to set
$z_0=\infty\in {\bf P}^1$,
but this is not a good choice as a base point since the point at
infinity is a singular point for the connection $\nabla .$  Instead of
fixing the holonomy of the connection $\nabla$ we will deform $\nabla^
0$ so
that certain monodromy matrices for a distinguished
fundamental solution $Y(z,a)$ to,
$$d_zY(z,a)-\Omega (z,a)Y(z,a)dz=0,$$
remain fixed as the points $a_1$ and $a_2$ vary in a neighborhood of
$a_1^0$ and $a_2^0$.  The distinguished fundamental solution will arise in
the course of fixing the behavior of the connection $\nabla$ in a
neighborhood of $\infty .$  Recall that our starting point in \S 1 was a
fundamental solution $F(z)$ to
$$d_zF-\Omega (z)Fdz=0,$$
where $\Omega (z)$ is given by (0.3).  We will fix the behavior of
$\nabla$ in a neighborhood of infinity by requiring that it be gauge
equivalent to the connection,
$$d_z-\Omega (z)dz,$$
in such a neighborhood, by a gauge transformation which is
$I+O(z^{-1})$ near infinity.  The fundamental solution $Y(z,a)$ will
be uniquely determined by the requirement that,
$$F(z)Y(z,a)^{-1}=I+O(z^{-1}),$$
for $z$ in a neighborhood of $\infty .$  The monodromy multipliers for
$Y(z,a)$ obtained by making counterclockwise circuits of $a_1$ and
$a_2$ will play the role of fixed holonomy.  Analytic continuation
of $Y(z,a)$ along a small counterclockwise circuit of $a_1$ will
transform $Y(z,a)$ into $Y(z,a)M$, where as in \S 1,
$$M=\left[\matrix{1&2\pi i\lambda\cr
0&1\cr}
\right].$$
Similarly, analytic continuation of $Y(z,a)$ in a small
counterclockwise circuit of $a_2$ will transform $Y(z,a)$ into
$Y(z,a)M^{-1}$.

We will next sketch the existence theory for the
deformation $\nabla$ by following the arguments in Malgrange [7].  In
fact, again following [7], it will be convenient to do a
little more.  It is possible to prove the existence of a
prolongation of the connection $\nabla^0$ to a connection in both
the $z$ and the $a$ variables in such a fashion as to incorporate
(1.15) as well as (1.14).

Let $D_j(r)$ denote the open disk of radius $r>0$ about the point $
a_j^0$ and write
$\Gamma_j(r)$  for the circle of radius $r$ about
the point $a_j^0$.  We suppose that for $j=1,2$ real numbers $\rho_{
1,j}$, $\rho_{2,j}$, and
$r_j$ are chosen so that
$$\rho_{2,j}>r_j>\rho_{1,j}>0$$
and the numbers $\rho_{2,j}$ are small enough so that the closed disks
$\bar {D}_1(\rho_{2,1})$ and $\bar {D}_2(\rho_{2,2})$ do not intersect.  We
write,
$$\eqalign{D_j&=D_j(\rho_{2,j}),\cr
\Gamma_j&=\Gamma_j(r_j),\cr
\delta D_j&=D_j\backslash\bar {D}_j(\rho_{1,j}).\cr}
$$
Choose $\epsilon >0$ so that
$\epsilon <\rho_{1,j}$ for $j=1,2$ and let $U$ denote the open ball of radius $
\epsilon$
about the point $(a_1,a_2)$ in ${\bf C}^2$.  Let
$$\Delta_j=\{(z,a)\in D_j\times U|z=a_j\}.$$
We pull back the connection one form $\Omega (z,a^0)dz$ to a connection
one form, $\Omega_jd(z-a_j)$ on $D_j\times U\backslash\Delta_j$ by the
translation
$$(z,a)\rightarrow z-a_j+a_j^0.$$
One finds
$$\Omega_j(z,a)=M(z-a_j+a_j^0,a^0)+\sum_k{{A_k(a^0)}\over {z-a_j+
a_j^0-a_k^0}},$$
where it will be convenient to introduce the notation
$$A_k(a^0)=(-1)^{k-1}\lambda C_k(a^0).$$
One can use this to define a connection
$$\nabla_j=d_z+d_a-\Omega_j(z,a)d(z-a_j)$$
on the trivial bundle
$$\left(D_j\times U\backslash\Delta_j\right)\times {\bf C}^2.$$
Observe that $\epsilon$ has been chosen small enough so that the only
singularities in $\Omega_j(z,a)$ for $z\in D_j$ occur when $z=a_j$ (where
there is a simple pole).  Let $\bar {D}$ denote the union of the closed
disks $\bar {D}_j(\rho_{1,j})$ for $j=1,2$.  Now we extend $\nabla^
0$ to a connection
on the trivial bundle
$${\bf C}\backslash\bar {D}\times U\times {\bf C}^2,$$
in the following fashion
$$\nabla_{\infty}=d_z+d_a-\Omega (z,a^0)dz.$$
Observe that $\nabla_{\infty}$ has no singularities on the domain we
consider it on.  Next we will show that the connections
$\nabla_j$ and $\nabla_{\infty}$ are gauge equivalent to one another
on the domain $\delta D_j\times U$ by a gauge
transformation $S_j(z,a)$ which is normalized by the condition
that $S_j(z,a^0)=I$ (the identity matrix).  More specifically
$$\nabla_j=S_j^{-1}\nabla_{\infty}S_j,\eqno (2.3)$$
on $\delta D_j\times U$.  To prove the existence of $S_j(z,a)$ we introduce
parallel transport relative to the connection $\nabla_j$.  If $\gamma$ is a
smooth curve in $\delta D_j\times U$ then we write $P_j(\gamma )$ for parallel
transport along the curve $\gamma$ relative to the connection $\nabla_
j$.
Since the bundle we consider is trivial we may think of
$P_j(\gamma )$ as a linear map on ${\bf C}^2.$  Now define
$$\eqalign{\gamma_j&(t,z,a)=(z,(1-t)a_1^0+ta_1,(1-t)a_2^0+ta_2)\cr}
\hbox{  for  }z\in D_j,\hbox{ }a\in U.$$
Then
$$[0,1]\ni t\rightarrow\gamma (t,z,a),$$
is a curve in $\delta D_j\times U$ which joins $(z,a^0)$ to $(z,a
)$.  We will
denote this curve by $\gamma_j(z,a)$ and define
$$S_j^{-1}(z,a)=P_j(\gamma_j(z,a)).$$
Observe that we have defined the inverse $S_j^{-1}$ rather than
$S_j$.
Because the connection $\nabla_j$ is integrable we can replace
$\gamma_j(z,a)$ by any smooth curve in $\delta D_j\times U$ which joins $
(z,a^0)$ to
$(z,a)$ but keeps the projection on the first coordinate (i.e., $
z)$
fixed.  The matrix $S_j$ of parallel translation thus satisfies
$$d_aS_j(z,a)=-S_j(z,a)\Omega_j(z,a)da.\eqno (2.4)$$
Now suppose that $z\in\delta D_j$ and let $B_j(z)$ be a small ball about
$z$ contained in the annulus $\delta D_j.$  Suppose that $u\in B_
j(z)$ and
let,
$$P_j(u,z|a),$$
denote parallel translation with respect to $\nabla_j$ along any smooth
curve that joins $(z,a)$ to $(u,a)$ whose projection on the first
coordinate stays inside $B_j(z)$ and whose projection on the
second coordinate is fixed at $a$.  Again because the connection
is integrable and the homotopy class of any such curve is
fixed,  parallel translation does not depend on the choice of such
a curve and we have
$$d_uP_j(u,z|a)=\Omega_j(u,a)P_j(u,z|a)du.\eqno (2.5)$$
Once again using the fact that parallel translation depends only
on the homotopy class of the path we can write
$$S_j(u,a)P_j(u,z|a)=P_j(u,z|a^0)S_j(z,a).$$
Differentiating this last relation with respect to $u$ using
(2.5) and then setting $u=z$ one finds
$$d_zS_j(z,a)+S_j(z,a)\Omega_j(z,a)dz=\Omega_j(z,a^0)S_j(z,a)dz.\eqno
(2.6)$$
But since $\Omega_j(z,a^0)=\Omega (z,a^0)$ we can combine (2.4) and (2.6) to
get (2.3).

Now we wish to factor $S_j(z,a),$
$$S_j(z,a)=\Sigma_{\infty}(z,a)^{-1}\Sigma_j(z,a),\eqno (2.7)$$
where $\Sigma_j(z,a)$ is a holomorphic, invertible matrix valued function
for $(z,a)\in D_j\times U$, and $\Sigma_{\infty}(z,a)$ is a holomorphic
invertible
matrix valued function for $z$ outside the union of the disks $\bar {
D}_j(\rho_{1,j})$
and $a\in U$.  We further require that $\Sigma_{\infty}(z,a)$ is normalized so
that
$$\Sigma_{\infty}(z,a)=I+O(z^{-1}).$$
In Malgrange [7] it is shown that by choosing a sufficiently
small neighborhood $U$ of $a^0$ this Wiener-Hopf factorization problem
can be solved.  Suppose now that $U$ is sufficiently small so
(2.7) is valid.  Combining (2.7) and (2.3) we find
$$\Sigma_j^{-1}\nabla_j\Sigma_j=\Sigma_{\infty}^{-1}\nabla_{\infty}
\Sigma_{\infty}\eqno (2.8)$$
over the set $\delta D_j\times U$ in the base.  On the left hand side of (2.8)
is a connection which extends to $D_j\times U\backslash\Delta_j$ with simple
poles
at $z=a_j$ and on the right hand side the connection extends to
a connection on ${\bf C}\backslash\bar {D}\times U$.  We denote by $
\nabla$ the connection whose
existence is thus assured by
(2.8).  Examining the singularity structure of this connection
at $z=a_j$ and $z=\infty$ using the properties of $\Sigma_j$ and $
\Sigma_{\infty}$ we find
that,
$$\nabla =d_z+d_a-M(z,a)dz-\sum_k{{A_k(a)}\over {z-a_k}}d(z-a_k),\eqno
(2.9)$$
where $M(z,a)$ is a polynomial in $z$ with the same leading terms
as $M(z,a^0)$ and hence the same leading terms as $\Omega (z)$ near infinity,
and $A_k(a)$ agrees with $A_k(a^0)$ at $a=a^0.$  To make contact with the
developments in \S 1 we introduce $C_k(a)$ by
$$A_k(a)=(-1)^{k-1}\lambda C_k(a).$$
The distinguished fundamental solution for $\nabla$ whose monodromy
is to remain fixed as a function of $a$ is given by,
$$Y(z,a)=\Sigma_{\infty}^{-1}(z,a)Y(z,a^0),$$
for $(z,a)\in {\bf C}\backslash\bar {D}\times U$.  It is clear that this has
exactly the same
monodromy multiplier as $Y(z,a^0)$, as $z$ makes a circuit of either
of the circles $\Gamma_j$.

Now we turn to the definition of the $\tau -\hbox{function}$.  Let $
H$ denote
the Hilbert space,
$$H=H^{{1\over 2}}(\Gamma_1)\oplus H^{{1\over 2}}(\Gamma_2),$$
where $H^{{1\over 2}}$ denotes the Sobolev space of order ${1\over
2}$.
Let $H_{+}$ denote the subspace of $H$ which consists of boundary
values of functions analytic in the open disks $D_j(r_j)$ for $j=
1,2$
(recall that the circle $\Gamma_j$ is the boundary of the disk $D_
j(r_j)$).  Let $H_{-}$ denote
the boundary values in $H$ of functions analytic in
$${\bf C}\backslash \{\bar {D}_1(r_1)\cup\bar {D}_2(r_2)\}$$
which vanish at $z=\infty$.  Then $H=H_{+}\oplus H_{-}$ and we denote the
projections relative to this decomposition by $P_{\pm}.$ If $S$ is a smooth
matrix
valued function defined on the circles $\Gamma_j$ for $j=1,2$ it defines a
map
$$S:H\rightarrow H$$
acting by multiplication by $S$ in the obvious way.  The Toeplitz
operator, $T_S,$ associated with $S$ is defined by
$$T_S=P_{+}SP_{+}.$$
We are now prepared to define the $\tau -\hbox{function}$.  Let $
S$ denote
the matrix valued function which is equal to $S_j$ on $\Gamma_j$ for
$j=1,2.$ Then, following Malgrange, we define
$$d\log\tau =\hbox{Tr}\left(T_S^{-1}T_{d_aS}-T_{S^{-1}d_aS}\right
).\eqno (2.10)$$
The inverse $T_S^{-1}$ is computable in terms of the factorization
(2.7) by the standard Wiener-Hopf technique and
(see lemma 6.7 in [7]) one finds
$$d\log\tau ={1\over {2\pi i}}\sum_j\int_{\Gamma_j}\hbox{Tr}\left
(\Sigma_j^{-1}d_z\Sigma_j\wedge\Sigma_{\infty}^{-1}d_a\Sigma_{\infty}\right
).\eqno (2.11)$$
In [10] the $\tau -\hbox{function}$ defined by (2.10) is shown to have the
following significance.  The family of subspaces
$$W(a)=Y(\cdot ,a^0)^{-1}S(\cdot ,a)H_{+}$$
determine boundary conditions for a Cauchy-Riemann operator
localized in the exterior of $\bar {D}$.  This Cauchy-Riemann operator
has a domain which incorporates functions which have suitable
branching at the points $a_j$.  The subspaces $W_{}(a)$ belong to
a Grassmannian over which there sits a holomorphic line
bundle $\hbox{det}^{*}$.  This line bundle has a connection whose
curvature vanishes over the family of subspaces $W(a)$.  The
$\tau -\hbox{function}$ defined by (2.10) is the ratio of the canonical
section of $\hbox{det}^{*}$ to a section of $\hbox{det}^{*}$ that is {\it flat}
with respect
to this connection.  It is also explained in [10] that this makes
it possible to regard $\tau$ $ $as the determinant of a singular
Cauchy-Riemann operator.  The reader is refered to [10] for
details.

{}From (2.8) and (2.9) we see that since
$$\Sigma_{\infty}^{-1}\nabla_{\infty}\Sigma_{\infty}=\nabla ,$$
we have
$$\Sigma_{\infty}^{-1}(z,a)d_a\Sigma_{\infty}(z,a)=-\sum_k{{A_k(a
)}\over {z-a_k}}da_k.\eqno (2.12)$$
Using (2.8) and (2.9) and
$$\Sigma_j^{-1}\nabla_j\Sigma_j=\nabla ,$$
one sees that
$$\Sigma_j^{-1}(z,a)d_z\Sigma_j(z,a)=\Omega (z,a)dz-\Sigma_j^{-1}
(z,a)\Omega_j(z,a)\Sigma_j(z,a)dz,\eqno (2.13)$$
where
$$\Omega (z,a)=M(z,a)+\sum_k{{A_k(a)}\over {z-a_k}}.$$
Since $\Sigma_j(z,a)$ is an invertible holomorphic function on $D_
j\times U$ it follows
that the pole type singularities in the two terms on the right
hand side of (2.13) must cancel.  One may easily confirm that
this implies
$$A_j(a)=\sigma_j^{-1}A_j(a^0)\sigma_j,\eqno (2.14)$$
where
$$\sigma_j:=\Sigma_j(a_j,a).$$
We write
$$\Sigma_j(z,a)=\sigma_j(I+\beta_j(z-a_j))+O\left((z-a_j)^2\right
)$$
for the first terms in the Taylor expansion of $\Sigma_j$ near
$z=a_j$.  Using (2.13) and (2.14) one finds that
$$\Sigma_j^{-1}\partial_z\Sigma_j=B_j(a)-\sigma_j^{-1}\left(B_j(a^
0)-[A_j(a^0),\beta_j]\right)\sigma_j+O\left((z-a_j)\right),\eqno
(2.15)$$
where
$$B_j(a)=M(a_j,a)+\sum_{k\ne j}{{A_k(a)}\over {a_j-a_k}}.$$
Substituting (2.15) and (2.12) into the formula (2.11) one finds
$$d_a\log\tau =\sum_j\hbox{Tr}\left(B_j(a)A_j(a)-B_j(a^0)A_j(a^0)\right
)da_j,\eqno (2.16)$$
where we made use the centrality of the trace, (2.14) again,
and the fact that
$$\hbox{Tr}\left([A_j(a^0),\beta_j]A_j(a^0)\right)=0.$$
If we use $A_j(a)=(-1)^{j-1}\lambda C_j(a)$ to compare (2.16) with (2.0) and
(2.2) we find
$$d_a\log\hbox{det}(1-\lambda K)=\sum_j\hbox{Tr}\left(B_j(a)A_j(a
)\right)da_j,$$
from which it follows that the $\tau -\hbox{function}$ differs from
$\hbox{det}(1-\lambda K)$ by the exponential of a linear function of $
a$.

\vskip.2in \par\noindent
\centerline{{\bf References}}
\vskip.2in \par\noindent
[1] Beals, R., Sattinger, D.H. : {\sl Integrable systems and isomonodromy
deformations, } preprint, Yale University.
\vskip.1in\par\noindent
[2] Harnad, J., Tracy, C.A., Widom, H. : {\sl Hamiltonian structure of
equations appearing in random matrices, } in {\it Low dimensional
topology and Quantum Field Theory}, ed. H.~Osborn, NATO ASI
Series B, Vol. {\bf 314}, Plenum Press (New York), 231--245 (1993).
\vskip.1in\par\noindent
[3] Its, A.R., Izergin, A.G., Korepin, V.E., Slavnov, N.A. : {\sl\ Differential
equations for quantum correlation functions, } Int. J. Mod. Physics B
{\bf 4}, 1003--1037 (1990).
\vskip.1in\par\noindent
[4] Jimbo, M., Miwa, T., M\^ori, Y., Sato, M. : {\sl Density matrix of an
impenetrable Bose gas and the fifth Painlev\'e transcendent, } Physica
{\bf 1D}, 80--158 (1980).
\vskip.1in\par\noindent
[5] Jimbo, M., Miwa, T., Ueno, K. : {\sl Monodromy preserving deformations of
linear ordinary differential equations with rational coefficients I.,}
Physica {\bf 2D}, 306--352 (1981).
\vskip.1in\par\noindent
[6] Jimbo, M., Miwa, T. : {\sl Monodromy preserving deformations of linear
ordinary differential equations with rational coefficients II, III.,} Physica
{\bf 2D}, 407--448 (1981); Physica {\bf 4D}, 26--46 (1983).
\vskip.1in\par\noindent
[7] Malgrange, B. : {\sl Sur les deformations isomonodromiques }, in
Mathematique et Physique: Seminaire de l'Ecole Normale
Superieure, 1979--1982 edited by L.B.~de Monvel, A.~Doudy, and
J.L.~Verdier, Birkhauser, Boston, 400--426 (1983).
\vskip.1in\par\noindent
[8] Mehta, M.L., : {\sl Random Matrices, } $2^{nd}$ edition, Academic Press,
San Diego.
\vskip.1in\par\noindent
[9] Moore, G. : {\sl Matrix models of 2D gravity and isomonodromic
deformation, } Prog. Theor. Physics Suppl. No. {\bf 102}, 255--285 (1990).
\vskip.1in\par\noindent
[10] Moore, G. : {\sl Geometry of the string equations, }Comm. Math. Phys.
{\bf 133}, 261--304 (1990).
\vskip.1in\par\noindent
[11] Palmer, J. : {\sl Determinants of Cauchy-Riemann operators as
$\tau -\hbox{functions}$, } Acta Applicandae Mathematicae {\bf 18} No. 3,
199--223
(1990).
\vskip.1in\par\noindent
[12] Tracy, C.A., Widom, H. : {\sl Introduction to random matrices, }
{{\it Geometric and Quantum Aspects of Integrable Systems }},
G.~F.~Helminck, Lecture Notes in Physics, Vol. {\bf 424},
Springer-Verlag (Berlin), 103--130, (1993).
\vskip.1in\par\noindent
[13] Tracy, C.A., Widom, H. : {\sl Level spacing distributions and the Airy
kernel, } to appear in Commun. Math. Phys.
\vskip.1in\par\noindent
[14] Tracy, C.A., Widom, H. : {\sl Level spacing distributions and the Bessel
kernel, } to appear in Commun. Math. Phys.
\vskip.1in\par\noindent
[15] Tracy, C.A., Widom, H. : {\sl Fredholm determinants, differential
equations and matrix models, } to appear in Commun. Math.
Phys.

\bye